# Spin orbit splitting of the photon induced Fano Resonance in an oscillating graphene electrostatic barrier

**by**

R. Biswas[1] and C. Sinha [2]

[1]Department of Physics, P.K.College, Contai, Purba Medinipur, W. B.- 721401, India.

[2] Department of Theoretical Physics, Indian Association for the Cultivation of Science,

Jadavpur, Kolkata-700032, India.

**Abstract:**

We investigate theoretically the effect of a time dependent oscillating potential on the transport property of the Dirac Fermion through a monolayer graphene electrostatic barrier under the influence of the Rashba spin orbit interaction. The time dependent problem is solved in the frame work of the non-perturbative Floquet approach. It is noted that the dynamic condition of the barrier may be controlled by tuning the Rashba parameter. Introduction of the spin orbit interaction causes splitting of the Fano resonance (FR), a characteristic feature in photon assisted tunneling. The separation between the spin split FR's gives an indirect measure of the fine structure of the quasi-hole bound state inside the barrier. The present findings on the Rashba splitting of the FR and its external control by tuning the oscillating field parameters might have potential for applications in spintronic devices, especially in the spin field effect transistors. The spin polarization of different Floquet sidebands is found to be quite sensitive to the spin-pseudospin interaction.

[*] Corresponding author. Tel.: 91 3220255020

E-mail: rbiswas.pkc@gmail.com

**I: Introduction. –** Spintronics[1, 2] is the branch of technology where the devices may be engineered by exploiting the spin degrees of freedom of the electrons for improved functionality. A spintronic device requires generation and manipulation of a spin polarized population of electrons, resulting in an excess of spin up or spin down charge states. The study of the spin dependent electronic transport opens up a new era in the field of information technology[3-8]. Similar to the digital information technology where the information are coded in terms of the 'ON' ('1') and 'OFF' ('0') states of the basic logic gates, the spin 'up' ('1') and spin 'down' ('0') states are expected to play the same role for encoding and manipulating information in spin controlled quantum information applications. The basic challenge for realization of such devices is to achieve high spin relaxation time and in this respect graphene may serve as a suitable platform with spin relaxation length of the order of micro-metre[9]. The very high current density and the rich spin dependent electronic properties[9-14], e.g., spin quantum Hall effect, long spin flip length, spin filtered edge states, adiabatic spin pumping, efficient spin injection etc. make graphene a promising candidate for the future nano-electronic and spintronic devices[15, 16]. One of the alternative ways to observe the spin dependent effect in a graphene (other than the application of the external magnetic field) is to introduce the spin orbit interaction (SOI) that plays a key role in controlling and manipulating the spin configuration and the spin current in practical devices. Graphene exhibits two types of SOI[10, 14-21]: (i) the intrinsic spin orbit interaction (ISOI) – that originates from the intra-atomic SOI of the carbon atoms and (ii) the extrinsic one, also known as the Rashba spin orbit interaction (RSOI) – that arises from the interaction between the graphene and a substrate or by the application of a gate voltage. Although the ISOI is quite small in graphene, it is capable to produce a band gap in the linear massless energy spectrum and mimics a quantum spin Hall insulator system with quantized spin Hall conductance[14]. On the other hand, the strength of the RSOI is externally tunable and is predicted[22, 23] to reach ~ 200meV at room temperature. This strong RSOI in graphene creates a considerable interest[18, 24, 25] to apply the RSOI in manipulating the spin polarization and spin relaxation effects in practical device applications. To generate spin polarized current, the RSOI acts in a way somewhat similar[26] to the case of the optical birefringence, the phenomenon of breaking up a polarized light in an optically anisotropic medium into two mutually perpendicular polarized lights moving with different velocities. For efficient spin polarization, the use of tunnel barrier turns out to be the common practice and as a result, the tunneling transport either through a static Rashba barrier[27-30] or through the electrostatic[31-34] or magnetic barriers[23] with static RSOI have been a subject of intense research in recent years. It may be mentioned here that the first evidence for the control of the electron spin precession by an external electric field via the SOI is the well known Datta - Das transistor[35]. Nowadays the study of SOI has drawn a major attention particularly because of its importance in spin relaxation, spin injection and spin transport in spintronic devices. Due to the relativistic effect, the RSOI depends explicitly on the linear momentum and the effective magnetic field produced by the Rashba spin orbit coupling is responsible for the precession of the electron spin for such a system using conventional semiconductors. In contrast, the RSOI in graphene is independent of the magnitude of the linear momentum of the quasi-particle and expresses the coupling between the real spin and the pseudo-spin due to the two sublattices in the honeycomb structure. Since in a graphene the direction of the pseudospin is either parallel or anti-parallel to the direction of the linear momentum, the RSOI in a graphene virtually arises due to the coupling between the direction of the linear momentum (not the magnitude) and the real spin of the Dirac fermions. This may be contrasted to the case of the conventional semiconductor (mentioned earlier). Another interesting feature of the SOI in graphene is that the inversion symmetry in the $k$ – space for the spin chiral states induced by the RSOI is destroyed[34].

Furthermore, the study of charge transport in a periodically driven tunneling system demands special interest particularly because, such a system could exhibit the Fano Resonance (FR)[36] that arises due to quantum interference between the continuum and the bound states and carries a signature of tunneling quasi-bound states. The Fano Resonance was also found earlier for the transmission of electrons through a dynamic quantum well both for the conventional heterostructure (with and without the SOI)[37-41] and the graphene[42] (without SOI) as well as through a graphene dynamic quantum barrier without the SOI[43]. The sinusoidally varying time dependent field of the dynamic quantum well or barrier causes photon induced bound-continuum transition of the electron with matching incident energy, leading to the appearance of the asymmetric FR in the transmission spectra. In fact, it was proposed[44] that the FR could be used as a probe of the phase coherence in the transport of electrons or other quasi-particles. Most of the earlier works on the spin dependent electronic transport in graphene considered the SOI under the static condition of the system. Although the spin dependent tunneling transport through a periodically driven system exists for the conventional heterostructures[38], no such work is available in the literature as yet for the cases of graphene quantum well or barrier structures. This motivated us to study the spin dependent transmission through a graphene electrostatic barrier under the periodically driven condition taking into account the effect of the Rashba spin orbit interaction. Here we have shown that the photon induced FR splits up due to the presence of the spin dependent interaction in the Hamiltonian and thereby confirms the spin splitting of the quasi-hole bound states inside the barrier. Although there is still no experimental observation of Rashba splitting in the prinstine graphene, the present study might create a boost in this direction.

**II: Theoretical Formulation. -** The model system considered here to study the spin dependent charge transport is shown in Fig. 1. Here the region – II with the RSOI having length $L$ is sandwiched between two regions – I and III without the RSOI. The region – II is subjected to a harmonically driven potential that could be realized by the application of a small AC signal voltage along with a static voltage induced by a local top gate. Neglecting the $KK'$ interaction, the Hamiltonian for the Dirac fermion in the region – II can be written as

$$H = H_0 + H_R + H_V = H^{/} + H_t \qquad (1)$$

where $H^{/}$ corresponds to the time independent part while $H_t$ refers to the time dependent part of the Hamiltonian. In eqn.(1) $H_0 = \mathbb{1}\otimes v_F(\vec{\sigma}\bullet\vec{p})$ , being the Hamiltonian in regions I and III describes the linear energy spectrum for the free Dirac fermion near the Dirac point, where the 2x2 identity matrix $\mathbb{1}$ in the kronecer product correspond to the real spin space and the latter part corresponds to the pseudospin space; $H_V = V_0 + V_1\cos(\omega t)$ for $0 < x < L$ and zero elsewhere, which gives the static barrier potential $V_0$ along with an AC component of amplitude $V_1$ and angular frequency $\omega$; $H_R = \frac{\lambda_R}{2}(\vec{s}\otimes\vec{\sigma})_Z$ for $0 < x < L$ and zero elsewhere, represents the RSOI term of strength $\lambda_R$, where $\vec{\sigma} \equiv (\sigma_x, \sigma_y)$ and $\vec{s} \equiv (s_x, s_y)$ are the Pauli matrices representing the pseudo-spin and the real spin of electrons respectively in the 2D honeycomb lattice of graphene. Here we assume that the temperature is low enough so that the electron-phonon interaction may be neglected. For the continuum description of the graphene in the Rashba region, we consider $L >> a$, $a$ being the lattice spacing. Since the spin dependent transmissions are identical for the two valleys, it is quite legitimate to consider the $K$ – valley only.

Since the Hamiltonian in eqn.(1) is independent of the y-coordinate, it preserves the translational symmetry in the y direction leading to the conservation of the momentum $(k_y)$. The corresponding wave function for the electron is therefore given by $\sim e^{ik_y y}$ and the energy dispersion relation for the Hamiltonian $H_0$ in eqn.(1) is $\varepsilon_p = \pm\sqrt{k_{px}^2 + k_y^2}$, with $p$ =1 and 2. From now onwards the wave vectors are expressed in units of $L_0^{-1}$, energy in $\hbar v_F/L_0$ and time in $L_0/v_F$. The corresponding forward moving (along the x-direction) plane waves may be written as $\xi_p(x) = N_p e^{ik_{px}x}$ ; $N_p$ denotes the four component spinor given by $N_p^T = A\left[\left(1, \frac{k_x+ik_y}{E}, 0{,}0\right)\delta_{1p} + \left(0, 0, 1, \frac{k_x+ik_y}{E}\right)\delta_{2p}\right]$ where $A$ is a constant, $T$ stands for the transpose and $\delta$ represents the delta function.

In the 4x4 notation the spin matrix can be represented by $\vec{s} = \boldsymbol{s}\otimes\mathbb{1}$, where $\boldsymbol{s}$'s are the 2x2 Pauli matrices in the real spin space. Then it can be shown that $\frac{\langle\xi_p|\vec{s}|\xi_p\rangle}{\langle\xi_p|\xi_p\rangle} = \frac{1}{2}\begin{pmatrix}0\\0\\S_p\end{pmatrix}$, where $s_p$ = ± 1 corresponding to $p$ = 1 and 2 respectively. The third component of the column matrix $\left(\frac{S_p}{2}\right)$ that represents the average value of the z-component of spin ($S_z$) corresponds to the spin up ($S_z$=1/2) or spin down ($S_z$=-1/2) state of the incident Dirac fermions. This indicates that the direction of real spin is perpendicular to the plane of the field free graphene sheet (the x-y plane) or in other words perpendicular to the direction of momentum. This is in contrast to the case of pseudo-spin of the Dirac fermion which is either parallel or anti-parallel to the direction of momentum.

Now the time independent part of the Hamiltonian in eqn.(1) including the static Rashba spin orbit interaction $(H^/ = H_0 + H_R + V_0)$ is written in a 4x4 notation as[34]

$$H^/ = \begin{pmatrix} V_0 & \left(-i\frac{\partial}{\partial x} - \frac{\partial}{\partial y}\right) & 0 & 0 \\ \left(-i\frac{\partial}{\partial x} + \frac{\partial}{\partial y}\right) & V_0 & i\lambda_R & 0 \\ 0 & -i\lambda_R & V_0 & \left(-i\frac{\partial}{\partial x} - \frac{\partial}{\partial y}\right) \\ 0 & 0 & \left(-i\frac{\partial}{\partial x} + \frac{\partial}{\partial y}\right) & V_0 \end{pmatrix} \qquad \text{............(2)}$$

The energy dispersion relation corresponding to the Hamiltonian $H^{/}$ in eqn.(2) is given by (following similar procedure as for $H_0$)

$$\varepsilon = V_0 + p\sqrt{k_{px}^{/2} + k_y^2 + \frac{\lambda_R^2}{4}} \pm \frac{\lambda_R}{2} \qquad (3)$$

while the corresponding solution representing the forward motion (along the x-direction) of the Dirac fermion with the wave vector $k_{px}^{/}$ may be written as $\psi_p(x) = M_p e^{ik_{px}^{/}x}$ with $M_p^T = A^{/}\left[\left(\frac{k_{px}^{/}-ik_y}{E-V}, 1, -i, -i\frac{k_{px}^{/}+ik_y}{E-V}\right)\delta_{1p} + \left(\frac{k_{px}^{/}-ik_y}{E-V}, 1, i, i\frac{k_{px}^{/}+ik_y}{E-V}\right)\delta_{2p}\right]$; $p = 1, 2$.

Evaluating the average value of the spin angular momentum in a similar manner (as for $H_0$) we obtain $\frac{\langle\psi_p|\vec{s}|\psi_p\rangle}{\langle\psi_p|\psi_p\rangle} = c\,\frac{s_p}{2}\begin{pmatrix}-k_y\\ k_{px}^{/}\\ 0\end{pmatrix}$; with $c = |E-V|\left(|E-V|^2 + k_{px}^{/2} + k_y^2\right)^{-1}$ and $s_p = +1\ or\ -1$ for $p$ = 1 or 2. In the limit of very small $\lambda_R$, the above equation reduces to[34] $\frac{\langle\psi_p|\vec{s}|\psi_p\rangle}{\langle\psi_p|\psi_p\rangle} = \frac{s_p}{2}\begin{pmatrix}-cos\theta_p\\ sin\theta_p\\ 0\end{pmatrix}$ with $\theta_p = tan^{-1}\left(\frac{k_{px}^{/}}{k_y}\right)$. It is thus clear that the average value of the z-component of spin angular momentum vanishes for the states $\psi_p$'s and the real spin of the Dirac fermions lies in the xy-plane. Therefore $k_{1x}^{/}$ and $k_{2x}^{/}$ may be interpreted as the wave vectors for two types of electrons having the z-component of spin in the opposite directions (in the plane of the graphene sheet) in presence of the external applied field.

Finally, in presence of both the spin orbit interaction and the time dependent potential the Floquet solution[45] corresponding to the full Hamiltonian (eqn.(1)), representing the propagating wave in the x-direction, can be written as

$$\psi_j(x,t) = \sum_{m,n=-\infty}^{\infty} A_m^j \left[\begin{pmatrix}\frac{p_+^m - ik_y}{E_m'}\\ 1\\ -i\\ \frac{-i(p_+^m + ik_y)}{E_m'}\end{pmatrix} e^{ip_+^m x}\,\delta_{j1} + \begin{pmatrix}\frac{p_-^m - ik_y}{E_m'}\\ 1\\ i\\ \frac{i(p_-^m + ik_y)}{E_m'}\end{pmatrix} e^{ip_-^m x}\,\delta_{j2}\right] J_{n-m}\left(\frac{V_1}{\omega}\right) e^{-i(E+n\omega)t} \qquad (4)$$

where $j$ = 1 or 2, $E'_m = E_m - V_0$ and $J_{n-m}$ represents the Bessel function of order $n$-$m$. The corresponding energy dispersion is obtained as $E_m = \pm\sqrt{\left(p_\pm^m\right)^2 + k_y^2}$, with $\left(p_\pm^m\right)^2 = \left[\left(E_m - V_0\right)\left(E_m - V_0 \pm \lambda_R\right) - k_y^2\right]$. Following the discussion for the time independent case one can conclude that under the influence of the time dependent potential the direction of spin angular momentum of the electrons are redistributed among the different sidebands.

To find the effect of the oscillating potential on the spin dependent tunneling transport, one has to find the solutions in the three different regions and then to apply the condition of continuity of wave functions at the two boundaries $x = 0$ and $x = L$ of the RSOI region. For an electron incident in the region – I characterized by the energy $E$ and the y-component of momentum $k_y$, the Floquet solution in three different regions corresponding to the Hamiltonian (eqn.(1)) can be written in a general form as follows:

$$\psi_r(x,y,t) = \sum_{m,n=-\infty}^{\infty}\left[A_m^r\begin{pmatrix}\frac{p_+^m - ik_y}{E'_m}\\ 1\\ i\\ \frac{i(p_+^m + ik_y)}{E'_m}\end{pmatrix}e^{ip_+^m x} + B_m^r\begin{pmatrix}\frac{-(p_+^m + ik_y)}{E'_m}\\ 1\\ i\\ \frac{-i(p_+^m - ik_y)}{E'_m}\end{pmatrix}e^{-ip_+^m x} + C_m^r\begin{pmatrix}\frac{p_-^m - ik_y}{E'_m}\\ 1\\ -i\\ \frac{-i(p_-^m + ik_y)}{E'_m}\end{pmatrix}e^{ip_-^m x} + D_m^r\begin{pmatrix}\frac{-(p_-^m + ik_y)}{E'_m}\\ 1\\ -i\\ \frac{i(p_-^m - ik_y)}{E'_m}\end{pmatrix}e^{-ik_-^m x}\right]$$

$$\text{X}\ \xi_{mn}\, e^{-i(E+n\omega)t}\, e^{ik_y y} \qquad (5)$$

where '$r$' corresponds to the regions I, II and III; $p_\pm^m = q_\pm^m$, $E'_m = E_m - V_0$ and $\xi_{mn} = J_{n-m}\left(\frac{V_1}{\omega}\right)$ for region – II and $p_\pm^m = k_\pm^m$, $E'_m = E_m$ and $\xi_{mn} = \delta_{mn}$ for regions – I and – III. Here $\left(q_\pm^m\right)^2 = \left[\left(E_m - V_0\right)\left(E_m - V_0 \pm 2\lambda_R\right) - k_y^2\right]$ and $\left(k_\pm^m\right)^2 = \left[E_m^2 - k_y^2\right]$ with $E_m = (E + m\omega)$. $J_{n-m}$ is the Bessel function for the first kind. The coefficients $A_m^r$, $B_m^r$, $C_m^r$, $D_m^r$ depend on the boundary conditions and the choice of the spin polarization in the incoming and the outgoing channels.

In order to derive the form of the transmission coefficients in the region III one has to consider the current density operators (in dimensionless unit) in the incident and transmitted channels in the x-direction written as

$$J^{in/tr}(x,t) = \psi^{*}_{in/tr}(x,t)\sigma_x\psi_{in/tr}(x,t) \qquad (6)$$

The above equation leads to the probability current density in the regions III for a particular (m-th) Floquet spin channel ($s^{/}$) corresponding to a given initial spin polarization ($s$) as

$$J^{m,tr}_{ss^{/}} = 4\frac{k^{m}_{s^{/}}}{E_m}\left|A^{m,tr}_{ss^{/}}\right|^2 \qquad (7)$$

where $A^{m,tr}_{ss^{/}}$ represents the amplitude of the m-th Floquet transmission channel. Finally equating the amplitude of the final outgoing wave to that of the incident wave, one can find the matrix $t^{ss'} = \left(A^{m,tr}_{ss^{/}} / A^{n,in}_{s}\right) = \left\{t^{ss'}_{mn}\right\}$, where $t^{ss'}_{mn}$ is the probability amplitude that an electron of spin $'s'$ incident with energy $E + n\omega$ is transmitted as an electron of spin $s'$ with energy $E + m\omega$ after crossing the Rashba barrier. Ultimately the transmission coefficients are written[46, 47] as $T_{ss'} = \sum_{m=-\infty}^{\infty} \frac{\cos\theta^{s^{/}}_{m}}{\cos\theta^{s}}\left|t^{ss'}_{m0}\right|^2$ with $\theta^{s^{/}}_{m} = \tan^{-1}\left(\frac{k_y}{k^{m}_{s^{/}}}\right)$, being the angle of transmission and $\theta^{s} = \tan^{-1}\left(\frac{k_y}{k^{0}_{s}}\right)$ being the incident angle.

Due to the presence of the SOI in the barrier region, the incident electron of a particular spin ('up' or 'down') undergoes spin flip, so that it can be transmitted in region-III with a finite probability both in the spin-up and spin-down states. So also is the case for reflection. Therefore the transmission probability for a particular spin injection $'s'$ can be expressed as a linear combination corresponding to the up and down spin states given by $T_s = T_{s\uparrow} + T_{s\downarrow}$. The total transmission probability and the z-component of the spin polarization for an un-polarized incident electron[27] are respectively given by $T = (1/2)(T_{\uparrow\uparrow} + T_{\uparrow\downarrow} + T_{\downarrow\uparrow} + T_{\downarrow\downarrow})$ and $P_z = (1/2)(T_{\uparrow\uparrow} + T_{\uparrow\downarrow} - T_{\downarrow\uparrow} - T_{\downarrow\downarrow})$.

**III. Results and Discussions**. - We now present the numerical results for the scattering of an incident electron by an oscillating spin-orbit coupled quantum barrier in a monolayer graphene. The RSOI is considered adiabatically so that the strength $\lambda_R$ does not change with the periodically varying time dependent electrostatic potential ($V_1 cos\omega t$). The energy of the incident electron ($E$), the static barrier height ($V_0$) and the width of the barrier ($L$) remain constant (e.g., 83 meV, 300 meV and 100 nm respectively) throughout the work unless otherwise specified. Fig. 2(a) displays the transmission probability ($T_s = T_{s\uparrow} + T_{s\downarrow}$) of the electron for normal incidence as a function of the Rashba parameter ($\lambda_R$) at four different amplitudes ($V_1$) of the oscillating potential (the solid line corresponds to the static case, i.e. $V_1$ = 0). It should be mentioned here that for the normal incidence, $T_s$ is found to be independent of the choice of '$s$' as was also noted by others[33]. The phenomenon of the Klein Tunneling[48] (unimpeded transmission through the quantum barrier) persists only for small values of $\lambda_R$ (up to ~ 20 meV) for all $V_1$, similar to the case without the SOI[46]. However, with the increase in $\lambda_R$, some oscillations set in the transmission profile as a result of the coupling between the pseudo-spin and the real spin of the Dirac Fermion, leading to the non-conservation of chirality for the system. Although the amplitude of oscillation increases with increasing $\lambda_R$, the average transmission decreases. In fact, in absence of the RSOI, no spin flip takes place inside the barrier and the incident spin state is transmitted in the collecting lead through the Rashba barrier without any change of the spin as well as without any attenuation due to the KT effect. But for $\lambda_R \neq 0$, spin flip might take place inside the barrier, similar to the case of the optical birefringence[26], so that each of the two spin polarization has a finite probability to be transmitted to the collecting lead. It may be mentioned in this context that in the case of light, incident normally on the optic axis of a doubly refracting crystal, it breaks up into two parts, one polarized along the optic axis while the other perpendicular to the former, both moving along the same direction but with different velocities similar to the case of Rashba barrier. However, in the former case (light), the two components recombine while leaving the crystal, unlike the present one that leads to a change in the state of polarization. In fact, in case of the Dirac fermions, due to quantum interference between the two parts (spin states) inside the barrier, some maxima and minima appear in the transmission profile for both the transmitted spin states depending on the strength of the pseudo-spin and real spin coupling (vide inset of Fig.2a). In the Rashba scale, the maxima of the spin up transmitted mode almost coincide with the minima of the spin down transmitted mode (i.e., $180^0$ phase lag) and vice versa. This indicates that 100% spin flip could be achieved for some distinct values of $\lambda_R$. The oscillating nature of the total transmission probability[33] ($T_s$) with increasing $\lambda_R$ could probably be attributed to the change of phase lag between the two transmitted spin states.

Regarding the effect of the external time dependent potential on the transmission coefficient ($T_s$), the following interesting features are noted from the Fig. 2(a). Distinct nodes and anti-nodes are found in the transmission profile (vide Fig. 2(a)) indicating that for certain values of $\lambda_R$ , the $T_s$ becomes insensitive (at the nodes) with respect to the amplitude variation of the time varying potential. In other words the time dependent barrier behaves as the static one. On the other hand, at the positions of the anti-nodes, the $T_s$ oscillates with maximum amplitudes with respect to the variation of $V_1$. These results might help to choose the optimal value of the parameter $V_1$ for controlling the current, since the amplitude at the anti-node is higher at higher values of $\lambda_R$.

For glancing incidence, on the other hand, an electron incident with a specific spin state (up or down) breaks up into two spin states (up and down) at the interface of the normal and the Rashba regions (i.e., electronic double refraction) and after that these two spin states move along different directions (unlike the case of normal incidence) with different group velocities, analogous to the optical case. Thus, for oblique incidence, $T_s$ changes in a different manner compared to that for the normal incidence. One such example is shown in Fig. 2(b) for glancing angle $\theta = 45^0$. The $T_s$ now depends on the choice of the incidence spin polarization and oscillates over the entire range of $\lambda_R$. For glancing incidence, no KT is observed and the $T_s$ exhibits a distinct competition between the two incident spin polarizations. By the application of the oscillating potential, the $T_s$ decreases systematically with increasing $V_1$ for both '$s$' at the lower range of $\lambda_R$, while the reverse is true for the higher range. This transition in the behavior of $T_s$ with respect to $V_1$ occurs at two different values of $\lambda_R$ depending on '$s$'. It is worth mentioning that $T_{\uparrow}(+\theta) = T_{\downarrow}(-\theta)$ and this angular symmetric feature maintains in the static and as well as in the dynamic conditions of the barrier, similar to the case of the field free Rashba barrier[27].

A completely different avenue opens up when the $T_s$ are plotted as a function of the incident electron energy ($E$), which is actually the most salient finding of the present work. Fig. 3 shows the energy dependent transmission spectrum for spin up electron incident on a weak Rashba barrier (small $\lambda_R$). The situation is almost identical to that of the oscillating electrostatic barrier without the SOI. In this weak limit of $\lambda_R$, $T_{\uparrow}(E) \approx T_{\downarrow}(E)$. A very sharp Fano type asymmetric resonance is noted in the tunneling spectrum (vide Fig. 3), not reported in the literature so far for the barrier transmission in graphene without the SOI. In fact, this type of asymmetric resonance is a characteristic for the photon induced transmission through the quantum well structure (for graphene see Lu et. al.[42] and for conventional heterostructure Zhang et. al.[38]). Recently we have shown[49] that tunneling through a bilayer graphene electrostatic barrier may also exhibit the asymmetric Fano resonance, that probably arises due to the quantum interference between the hole quasi bound state and the continuum. The present FR in Fig. 3 occurs due to the quantum interference between the hole quasi-bound state inside the barrier and the hole continuum via photon exchange. With the increase in $V_1$ (the amplitude of the oscillating potential), the width of the FR increases while the position of the FR remains unaltered, the latter being sharply dependent on the frequency of the oscillating potential (not shown).

A new feature may be noted in the transmission spectra when the spin - pseudospin coupling strength becomes appreciable. Fig. 4 shows such an energy dependent transmission for a spin up incident Dirac Fermion for three different values of $V_1$. Unlike the case of zero (or very small) $\lambda_R$, two closely spaced FR's are noted in Fig.4. The appearance of the double Fano resonances (DFR) indicates that the single (SOI free) FR splits into two FR's due to the inclusion of RSOI. The origin of the spin orbit splitting of the FR could be understood as follows: In presence of the RSOI, the spin degeneracy of the quasi hole bound state inside the barrier is removed, resulting in the appearance of two quasi-hole bound states. The quantum interference between these two discrete hole states with the hole continuum (inside the barrier) via photon exchange leads to the fragmentation of the single FR into two FR's at two different incident energies for a particular frequency of the oscillating field. Thus we find that the two fold spin symmetry of the quasi-bound state is broken by the RSOI, indicating that the spin and the pseudo-spin are not independent degrees of freedom. To our knowledge, the present spin dependent DFR is the first finding in the literature. The nature of the DFR is sharply dependent on the amplitude $V_1$, as is noted from the inset of Fig. 4. It is interesting to note that, the ratio of the $T_s$ at the peak to that of the dip (PDR) are different for the two spin split FR's. At intermediate $V_1$ (~ 5 meV) the DFR combines to form a single nearly symmetric sharp resonance, somewhat analogous to the case of Fano-Feshbach resonance[50] (the appearance of a constructive interference due to overlap of two closely spaced FRs). Finally, at higher amplitude, the DFR disappears leaving a single FR of extended width as depicted in Fig. 4. This indicates the phenomenon of collapse and revival of the spin split hole quasi bound state inside the barrier. Thus by controlling the amplitude of the oscillating potential, it is possible to switch from double to single Fano picture in a graphene Rashba barrier.

In order to study the effect of the incident spin state on the spin split DFR, we plot in Fig.5 the $T_s$ for both $s = \uparrow \,\&\, \downarrow$ with three different Rashba parameters. Two important points may be noted from Fig. 5 as follows. First, on introduction of the RSOI with appreciable strength, the single FR splits into two with different PDR and the separation between the FR's increases with increasing $\lambda_R$. Second, the effect of the incident spin state becomes prominent only at appreciable value of $\lambda_R$ and the two FR's (marked by their different PDR) interestingly exchange their places on changing the incident spin state.

Next we present the spin polarization of electron transmission through the monolayer graphene electrostatic barrier. Fig. 6 displays the z-component of the of spin polarization of transmission ($P_z$) as a function of incident angle for different values of the Rashba parameter in case of a static electrostatic barrier. The spin polarization of electron through a field free static Rashba barrier was also studied earlier by Bercioux *et. al.*[27] where it was reported that, a finite spin polarization may be achieved only when both the intrinsic and Rashba spin orbit interactions are simultaneously present. In contrast, the present Fig. 6 indicates that even in absence of the intrinsic spin orbit coupling (ISOC), an electrostatic barrier is capable to produce finite $P_z$ in presence of the RSOI. It is probably due to the fact that the external electrostatic potential causes the spin to have a component parallel to the electron motion[34] and thereby results in a finite spin polarization in the transmitted channel. For normal incidence, $P_z$ is zero for all $\lambda_R$ which may be attributed to the cloaking effect of the Klein tunneling that preserves the incident spin polarization during the transmission process. The $P_z$ remains an odd function of the incident angle, similar to the case of $V_0 = 0$. For finite $V_0$, the $P_z$ is only appreciable around the intermediate glancing incidence (from $30^0$ to $70^0$) of the electron and the change in the strength of the RSOI causes oscillation in $P_z$. This oscillating behavior arises due to the phase coherence of the electron wave function during the tunneling process. For an unpolarised electron incidence, the maximum probability of spin down (the direction normally downwards to the graphene plane) transmission is greater than that for the spin up transmission for all values of $\lambda_R$ (vide Fig. 6).

The signature of the time varying electrostatic potential is well reflected from Fig. 7, where the angular dependence of $P_z$ is displayed for an oscillating barrier with appreciable value of the Rashba parameter. It is worth mentioning that in reality a very high RSOI strength can be achieved[51] by covalent bonding with absorbates. In general the magnitude of the spin polarization is quenched by the application of the time varying potential, particularly at lower strength of the coupling. However in some angular ranges, (e.g. for F = 3, vide Fig. 7) the magnitude of $P_z$ may exceed the field free value, particularly at higher value of $\lambda_R$. The characteristic FR is noted in the polarization profile, especially at higher angular incidence on the periodically driven electrostatic barrier. The position and nature of the FR's are sharply dependant on the strength and frequency of the oscillating potential. The sudden reversal of $P_z$ (the characteristic of the FR) around the FR would make such structures suitable for the spintronics devices.

Finally, to study the nature of the spin polarization in different Floquet sidebands, Fig. 8 (a) displays the z-component ($P_z^{0,\pm1}$) of the sideband spin polarization (superscript '0' represents the central and +1 (-1) represents the first absorption (emission) side band) against the strength of the spin orbit interaction $\lambda_R$ for a fixed small value of $k_y$ (=1.16). It may be noted from Fig. 8a that similar to the case of overall polarization ($P_z$), the side band polarizations are also oscillatory in nature. For small values of $k_y$ the magnitude of the spin polarization is very small (< 1%) for all the sidebands which is probably due to the effect of the Klein transmission (the phenomenon for the normal and near normal incidence). Although the $P_z^0$ is greater than the $P_z^{\pm 1}$, particularly at higher $\lambda_R$, the former is lower than $P_z^{\pm 2}$ for all $\lambda_R$. The sideband polarization is higher for the absorption band (vide Fig. 8b) than for the emission one in the case of even order. While the reverse is true for the odd ones, particularly at the lower values of $\lambda_R$. This leads to the conclusion that the two photon processes are more spin polarized than the single photon one.

For higher glancing incidence, a different situation arises in the nature of the sideband spin polarization displayed in Figs. 9(a) and 9(b). At higher value of $k_y$, the tunneling transmission is forbidden via the process of photon emission (to maintain the conservation of energy and the y-component of momentum) and the allowed transmission occurs with a comparatively higher spin polarization (~10 – 20 %, vide Fig. 9a) than that for the lower values of $k_y$. In this case (for higher $k_y$) both the $P_z^0$ and the $P_z^{+1}$ exhibit two resonant peaks (representing spin polarization in the upward direction) at the same values of $\lambda_R$ and their comparative magnitudes ($|P_z^{0,+1,+2}|$) decrease with the increase in the side band order. It should be mentioned here that in the field free situation, the spin polarization is almost zero for the parameter used in Fig. 9(a). On the other hand, the polarization is strongly suppressed (vide Fig. 9b) as compared to the field free one by the application of the time dependent potential, for higher range of $\lambda_R$ (vide inset of Fig. 9b). The relative magnitudes of $P_z$ in the Floquet bands are sharply dependent on the strength of the spin - pseudospin interaction.

**IV. Summary. -** In conclusion, the spin dependent charge transport in a monolayer graphene oscillating electrostatic barrier is studied. The most salient features of the study are as follows. For normal incidence, the Rashba barrier is unable to distinguish between the incident spin states of the charge carriers. The finite probability of transmission at the collecting lead for both spin-up and spin-down charge carriers clearly reveals the occurrence of spin flip inside the Rashba barrier due to the spin - pseudospin interaction. It is noted for the first time that a barrier structure may also exhibit the photon induced Fano resonance (like the quantum well structure). The effect of spin of the charge carrier causes the fragmentation of the Fano resonance resulting in the removal of spin degeneracy of the quasi hole bound states inside the barrier. This feature of the spin split Fano spectrum may be tuned by controlling the Rashba parameter or by changing the frequency and amplitude of the oscillating potential. The creation of the two spin split tunneling hole sub-bands and their control through the external parameters might explore the system for spintronic applications e.g., in the realization of spin qubits, in graphene based memory devices, quantum computations etc.. Regarding the spin polarization, three important points may be noted: First, for normal incidence, the transmitted electrons remain unpolarized due to the effect of the Klein tunneling. Second, for glancing incidence, the polarization is generally suppressed by the application of the oscillating potential. Third, at low grazing incidence, the electron transmission via two photon processes is more spin polarized than through the single or no photon processes while the reverse is noted for higher grazing incidence. Finally, the present findings might have potential importance for the applications in spintronics for graphene based systems.

**V. References:**

[1]S. A. Wolf, et. al. Science **294**, 1488 (2001).

[2]I. Zutic, J. Fabian and S. Das Sharma, Rev. Mod. Phys. **76**, 323 (2004).

[3]K. C. Nowack, F. H. L. Koppens, Y. V. Nazarov, and L. M. K. Vandersypen, Science **318**, 1430 (2007).

[4]J. M. Elzerman, R. Hanson, L. H. Willems van Beveren, B. Witkamp, L. M. K. Vandersypen and L. P. Kouwenhoven, Nature **430**, 431 (2004).

[5]J. A. Petta, Science **309**, 2180 (2005).

[6]A. G. Burkard, Phys. Rev. B **59**, 2070 (1999).

[7]P. Recher and B. Trauzettel, Nanotechnology **21**, 302001 (2010).

[8]D. D. Awschalom and M. E. Flatte, Nat. Phys. **3**, 153 (2007).

[9]N. Tombros, C. Jozsa, M. Popinciuc, H. T. Jonkman and B. J. van Wees, Nature **448**, 571 (2007) .

[10]A. H. Castro Neto, F. Guinea, N. M. R. Peres, K. S. Novoselov and A. K. Geim, Rev. Mod. Phys., **81**, 109 (2009).

[11]S. Cho, Y. F. Chen and M. S. Fuhrer, Appl. Phys. Lett. **91**, 123105 (2007).

[12]M. Nishioka and A. M. Goldman, Appl. Phys. Lett. **90**, 252505 (2007).

[13]W. H. Wang, et. al. Phys. Rev. B **77**, 020402 (2008).

[14]C. L. Kane and E. J. Male, Phys. Rev. Lett. **95**, 226801 (2005).

[15]B. Trauzettel, D. V. Bulaev, D. Loss and G. Burkard, Nat. Phys. **3**, 192 (2007).

[16]D. Pesin and A. H. C. MacDonald, Nat. Mat. **11**, 409 (2012).

[17]H. Min, J. E. Hill, N. Sinitsyn, B. Sahu, L. Kleinman and A. H. C. MacDonald, *Phys. Rev.* B **74**, 165310 **(**2006**)**.

[18]E. I. Rashba, Phys. Rev. B **79**, 161409 (2009).

[19]A. H. Castro Neto and F. Guinea, Phys. Rev. Lett. **103**, 026804 (2009).

[20]D. Huertas-Hernando, F. Guinea and A. Brataas, Phys. Rev. B **74**, 155426 (2006).

[21]A. Dyrdal, V. K. Dugaev and J. Barnas, Phys. Rev. B **80**, 155444 (2009).

[22]Y. S. Dedkov, M. Fonin, U. Ru¨diger and C. Laubschat, Phys. Rev. Lett. **100**, 107602 (2008).

[23]C. Bai, J. Wang, S. Jia, and Y. Yang, Appl. Phys. Lett. **96**, 223102 (2010).

[24]C. Ertler, S. Konschuh, M. Gmitra and J. Fabian, Phys. Rev. B **80**, 041405 (2009).

[25]D. Huertas-Hernando, F. Guinea and A. Brataas, Phys. Rev. Lett. **103**, 146801 (2009).

[26]M. M. Asmar and S. E. Ulloa, Phys. Rev. B 87, 075420(2013).

[27]D. Bercioux and A. D. Mertino, Phys. Rev. B **81**, 165410 (2010).

[28]L. Lenz and D. Bercioux, Europhys. Lett. **96**, 27006 (2011).

[29]Q. Zhang, Z. Lin and K. S. Chan, Appl. Phys. Lett. **102**, 142407 (2013).

[30]C. Bai, J. Wang, J. Tian and Y. Yang, Physica E **43**, 207 (2010).

[31]A. Yamakage, K. I. Imura, J. Cayssol, and Y. Kuramoto, Europhys. Lett. **87**, 47005(2009).

[32]M. H. Liu, J. Bundesmann and K. Richter, Phys. Rev. B **85**, 085406(2012).

[33]E. Faizabadi and F. Sattari, J. Appl. Phys. **111**, 093724 (2012).

[34]Kh. Shakouri, M. R. Masir, A. Jellal, E. B. Choubabi and F. M. Peeters, Phys. Rev. B 88 115408 (2013).

[35]S. Dutta and B. Das, Appl. Phys. Lett. **56**, 665 (1990).

[36]U. Fano, Phys. Rev. **124**, 1866 (1961).

[37]M. Buttiker, Phys. Rev. Lett. **49**, 1739 (1982).

[38]C. X. Zhang, Y. H. Nie and J. Q. Liang, Phys. Rev. B **73**, 085307 (2006).

[39]M. Wagner, Phys. Rev. Lett. **76**, 4010 (1996).

[40]W. Li and L. E. Reichl, Phys. Rev. B **60**, 15732 (1999).

[41]I. A. Shelykh, Phys. Rev. B **70,** 205328 (2004).

[42]W. T. Lu, S. J. Wang, W. Li, Y. L. Wang, C. Z. Te and H. Jiang, J. Appl. Phys. **111,** 103717 (2012).

[43]R. Biswas and C. Sinha, J. Appl. Phys. **114,** 183706 (2013).

[44]S. J. Xu, S. J. Xiong, J. Liu and H. Z. Zheng, Europhys. Lett. **75**, 875 (2006).
[45]J. H. Shirley, Phys. Rev. **138**, B979 (1965).
[46]C. Sinha and R. Biswas, Appl. Phys. Lett. **100,** 183107 (2012).
[47]M. I. Katsnelson, K. S. Novoselov and A. K. Geim, Nat. Phys. **2,** 620 (2006).
[48]M. A. Zeb, K. Sabeeh and M. Tahir, Phys. Rev. B **78**, 165420 (2008).
[49]C. Sinha and R. Biswas, Phys. Rev. B **84,** 155439 (2011).
[50]A. E. Miroshnichenko, Phys. Rev. E **79,** 026611 (2009).
[51]D. Ma, Z. Li and Z. Yang, Carbon **50**, 297 (2012)

**Figure Captions**:

Fig.1: (a) Schematic diagram for a time periodically modulated graphene electrostatic barrier of height $V_0$ and length $L$ subjected to Rashba Spin orbit interaction. (b) Electron birefringence in Rashba region for glancing incidence of a spin polarized electron.

Fig. 2(a): Transmission coefficient ($T_s = T_{s\uparrow} + T_{s\downarrow}$, in the present case $T_\uparrow = T_\downarrow$) plotted as a function of Rashba parameter $\lambda_R$ (in meV) for $L$= 100 nm, $\theta = 0^0$, $E$ = 83 meV, $V_0$= 300 meV and $\omega$ = 11 meV. $V_1$= 0, Solid line (black); 5 meV, dash line (red); 10 meV, dot line (blue); 20 meV, dash-dot line (purple). Inset: Solid (black), $T_{\uparrow\uparrow}$ for $V_1$ = 0; Dot (black), $T_{\uparrow\uparrow}$ for $V_1$ = 20 meV; Dash (red), $T_{\uparrow\downarrow}$ for $V_1$ = 0; Dash-Dot (black), $T_{\uparrow\downarrow}$ for $V_1$ = 20 meV. (b) Same as (a) but for $\theta$= $45^0$. Solid and dash lines for spin up ($T_\uparrow$) and spin down ($T_\downarrow$) incidence respectively. $V_1$= 0 (black), = 5 meV (blue) and = 10 meV (red). (Color online only).

Fig. 3: Transmission coefficient ($T_s = T_\uparrow \approx T_\downarrow$) plotted against incident energy $E$ (in meV) for $L$= 100 nm, $V_0$= 300 meV, $\lambda_R$ = 1 meV, $k_y$ = 0.012 nm$^{-1}$ and $\omega$ = 8 meV . $V_1$= 0, Solid (black) line; 10 meV, dash (red) line; 15 meV, dot (blue) line. (Color online only).

Fig. 4: Transmission coefficient ($T_s$ = $T_\uparrow = T_{\uparrow\uparrow} + T_{\uparrow\downarrow}$) plotted against incident energy $E$ for $\lambda_R$ = 10 meV. Others parameters are same as Fig. 3. (Color online only).

Fig. 5: $T_s$ verses $E$ at $V_1$ = 1 meV. Solid (black) line for $T_\uparrow$ (spin up incidence) and Dash (black) line for $T_\downarrow$ (for spin down incidence) for $\lambda_R$ = 10 meV; Dash-dot (red) line for $T_\uparrow$ (spin up incidence) and Dot (red) line for $T_\downarrow$ (for spin down incidence) for $\lambda_R$ = 1 meV; Dash-dot-dot (blue) line for $T_\uparrow = T_\downarrow$ for $\lambda_R$ = 0 meV. (Color online only).

Fig. 6: The z-component of polarization ($P_z$) plotted as a function of the incident angle for $L$= 100 nm, $V_0$= 300 meV, $E$ = 83 meV and $V_1$ = 0 meV. Here $\lambda_R$ = 8 meV (Solid-black), 16 meV (dash-red), 24 meV (dot-blue), 32 meV (dash-dot-green) and 40 meV (dash-dot-dot-purple).

Fig. 7: Same as Fig. 6 but for $\lambda_R$ = 128 meV, ω = 8 meV. Here $V_1$ = 0 meV (Solid-black), 8 meV (dash-red) and 24 meV (dot-blue). Inset: For $\lambda_R$ = 16 meV; for $V_1$ = 0 meV (dotted line) and 8 meV (solid line).

Fig. 8: The z-component of the sideband polarization ($P_z$ ) as a function of the Rashba parameter $\lambda_R$ for ω = 4 meV, $V_1$ = 20 meV and $k_y$ = 0.014 nm$^{-1}$. Other parameters are same as Fig. 6. Here (a) Solid-black line, $P_z$ for the first absorption band (i.e., $P_z^{+1}$), dotted-red line, $P_z$ for the central band (i.e., $P_z^0$) and dash-blue line, $P_z$ for the first emission band (i.e., $P_z^{-1}$). (b) Solid-black line, $P_z$ for the first absorption band (i.e., $P_z^{+2}$), dotted-red line, $P_z$ for the central band (i.e., $P_z^0$) and dash-blue line, $P_z$ for the first emission band (i.e., $P_z^{-2}$).

Fig. 9: Same as Fig. 8 but for $k_y$ = 0.125 nm$^{-1}$. (a) dash-black line for $P_z^{+2}$, solid-red line for $P_z^{+1}$ and dotted-blue line for $P_z^0$ . (b) solid-black line for $P_z^{+2}$, dotted-red line for $P_z^{+1}$ and dash-blue line for $P_z^0$.

a)

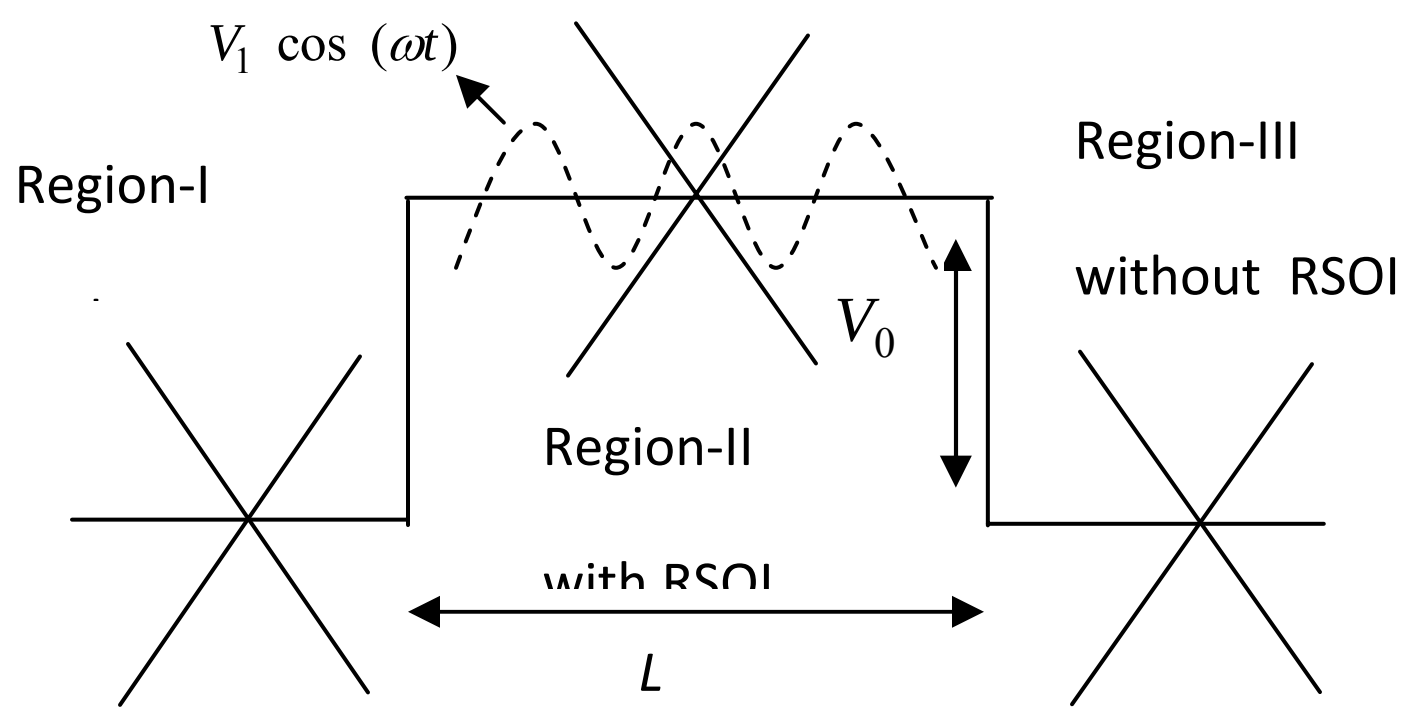

(b)

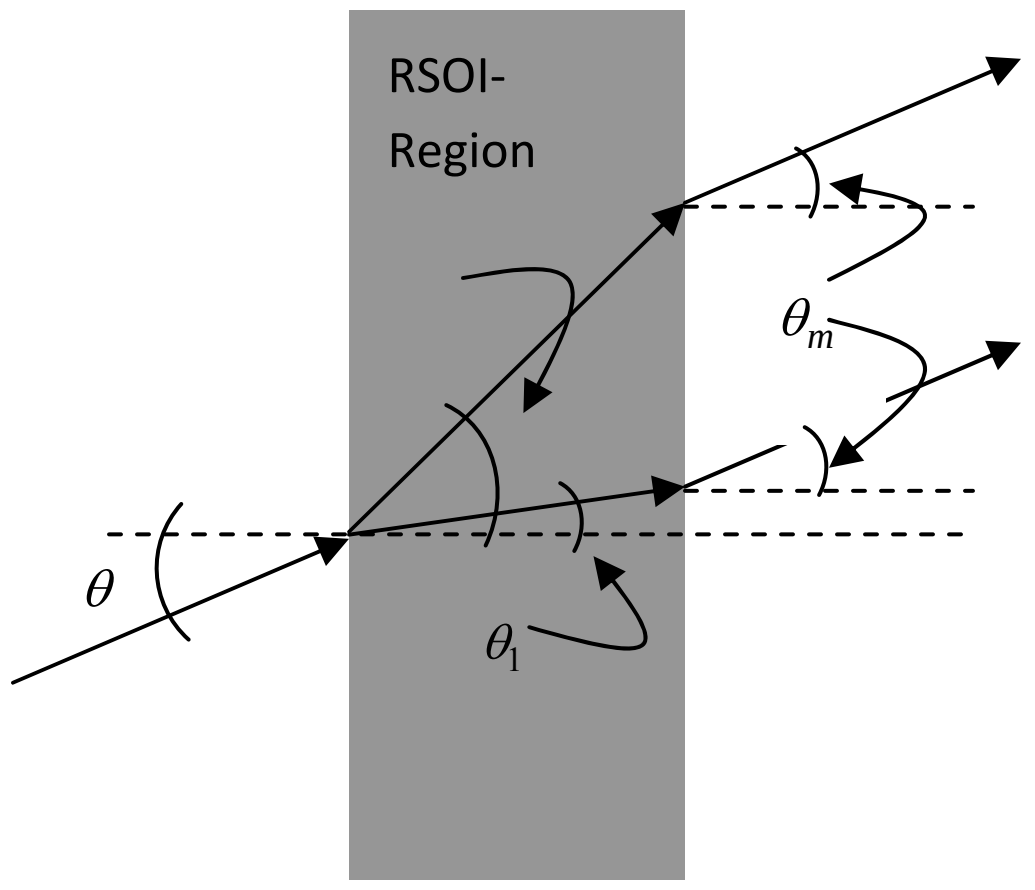
RSOI-
Region
$\theta_m$
$\theta$
$\theta_1$

Fig. 1

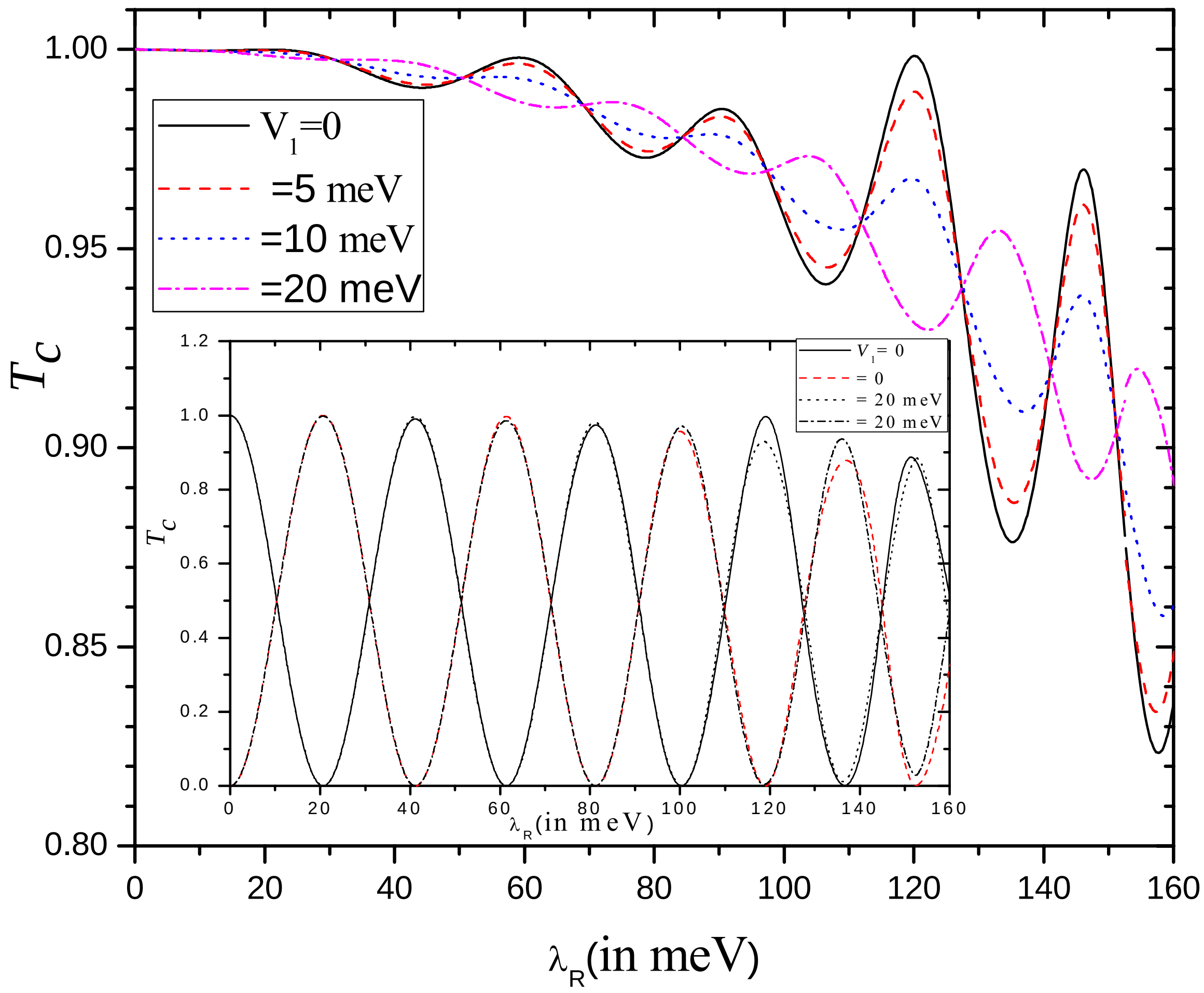

Fig.2a

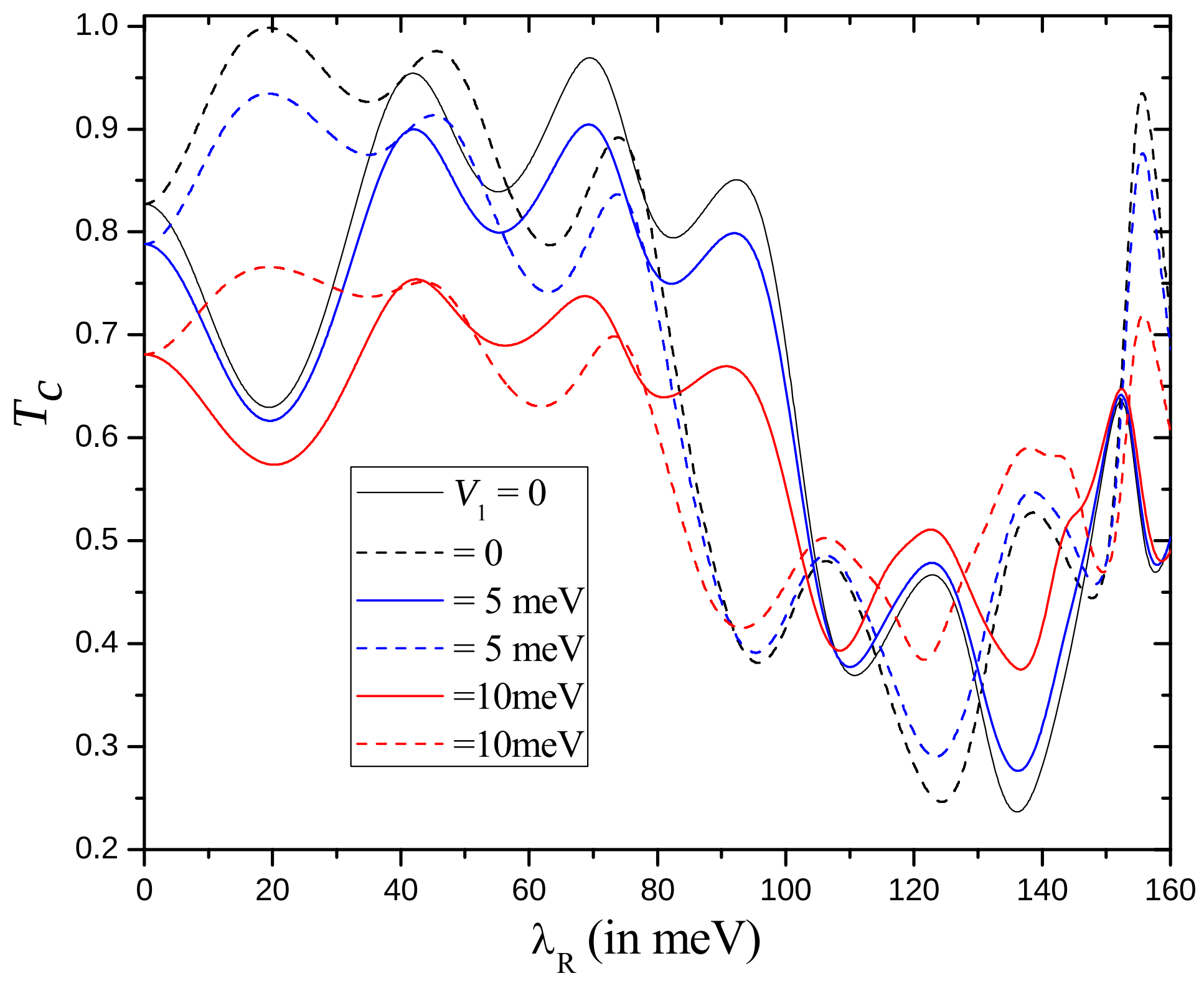

Fig.2b

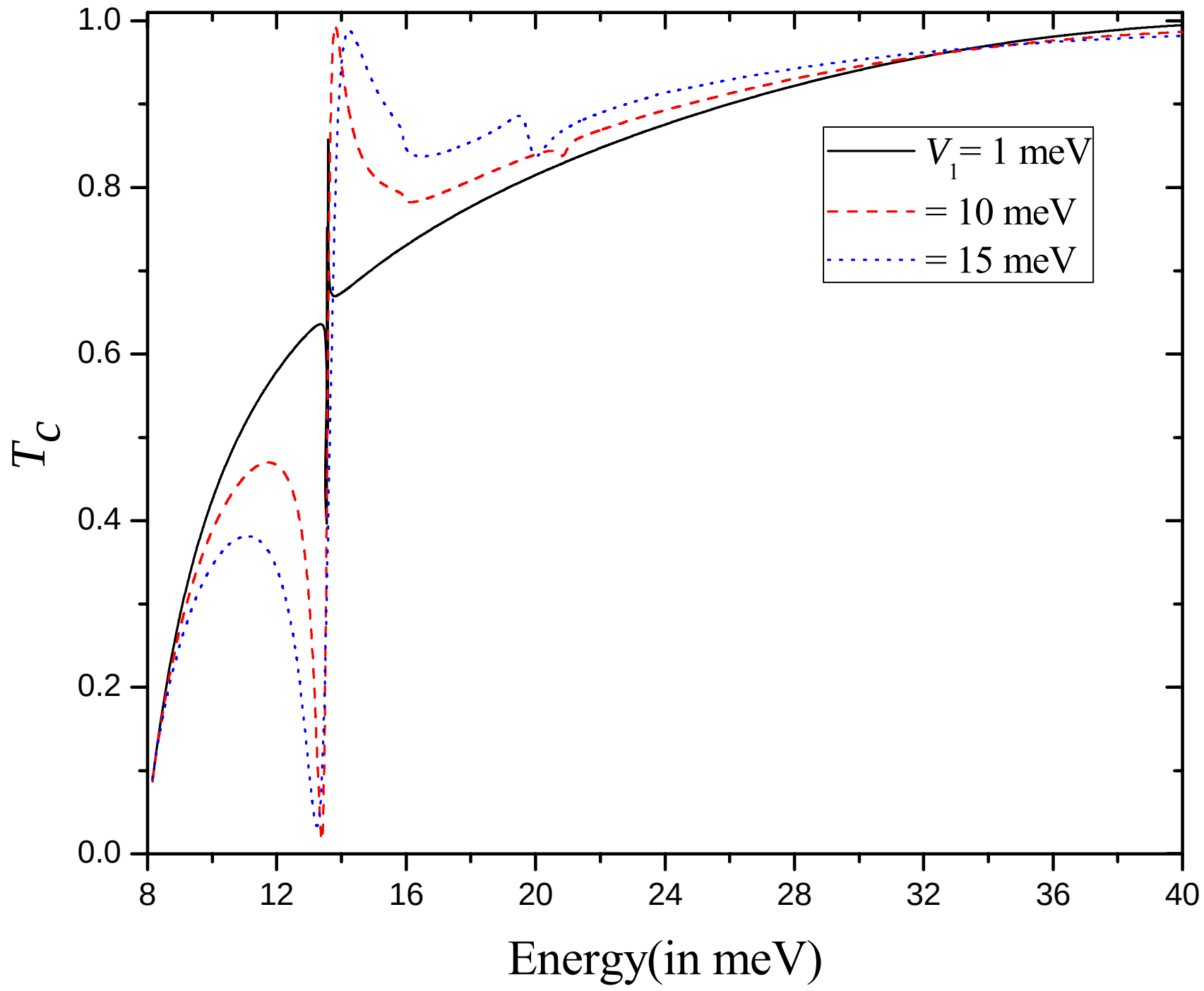

Fig.3

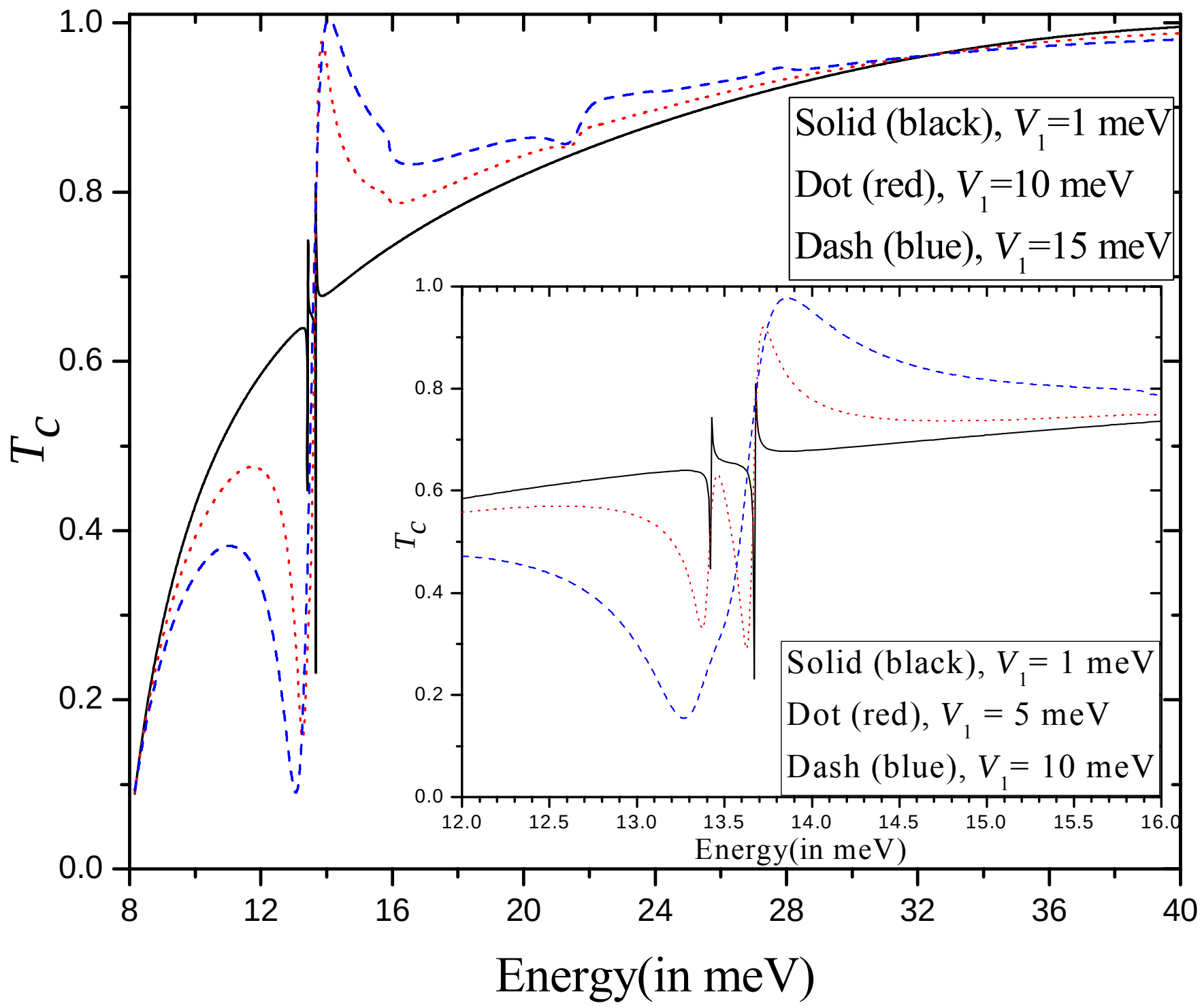

Fig.4

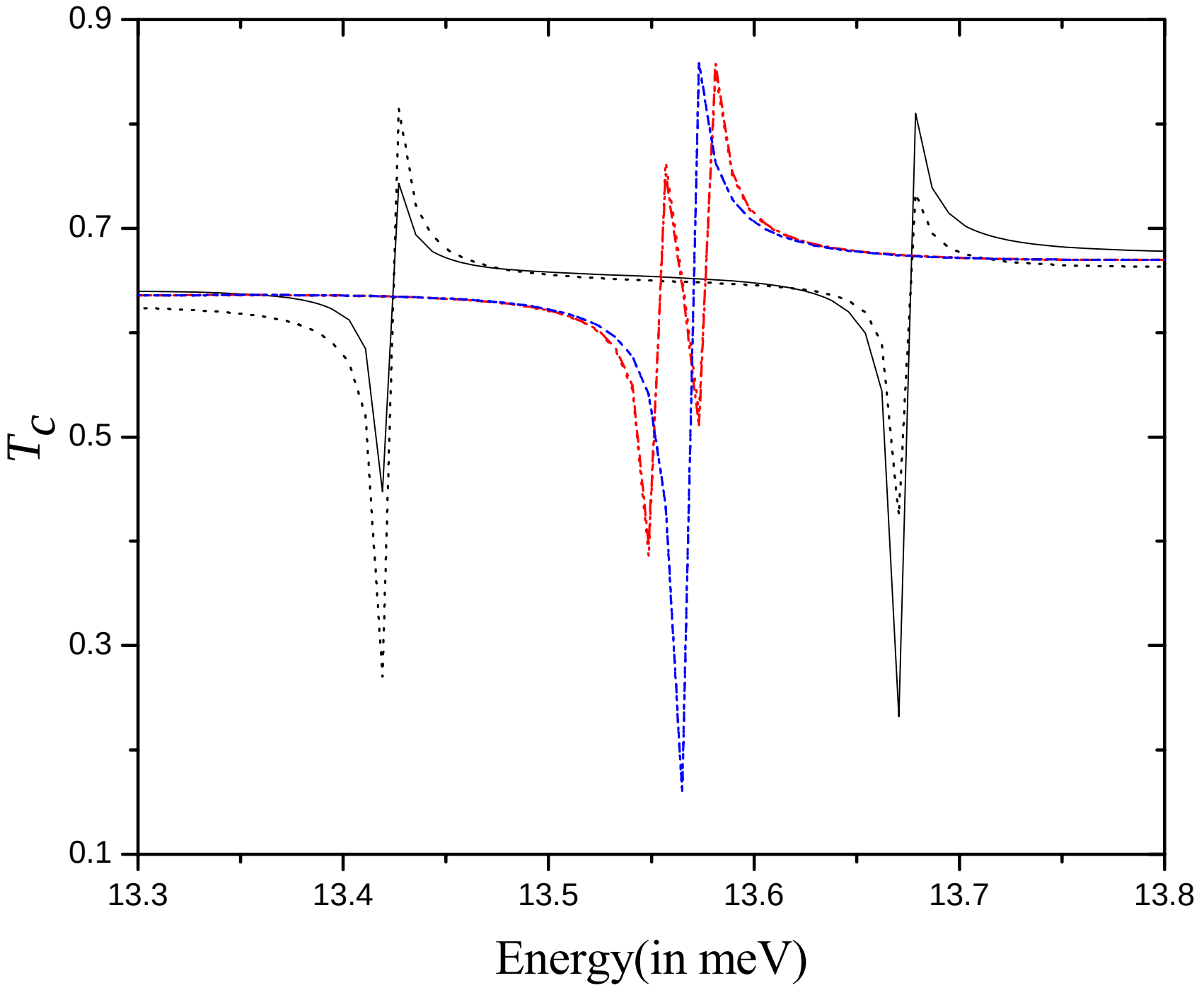

Fig.5

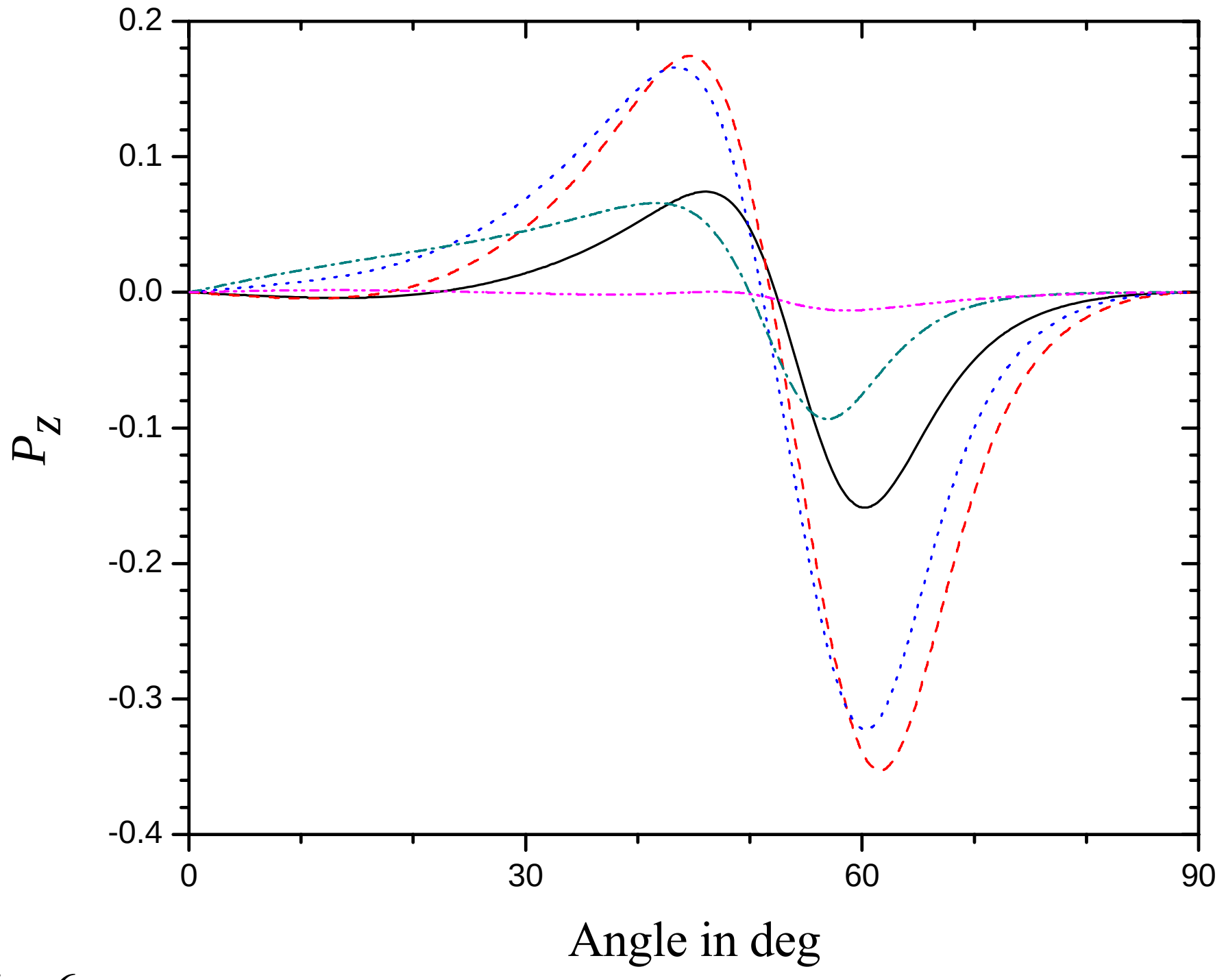

Fig. 6

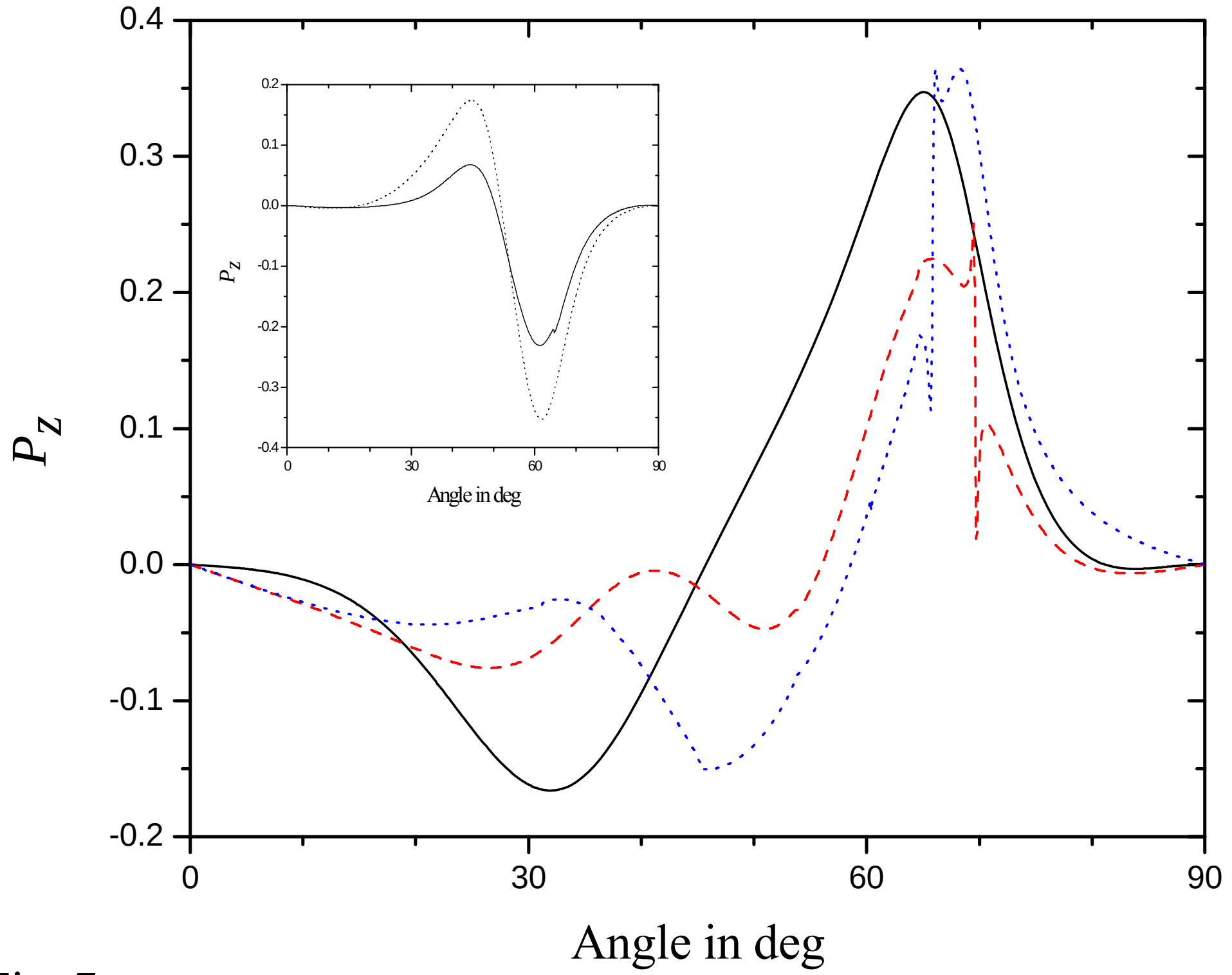

Fig. 7

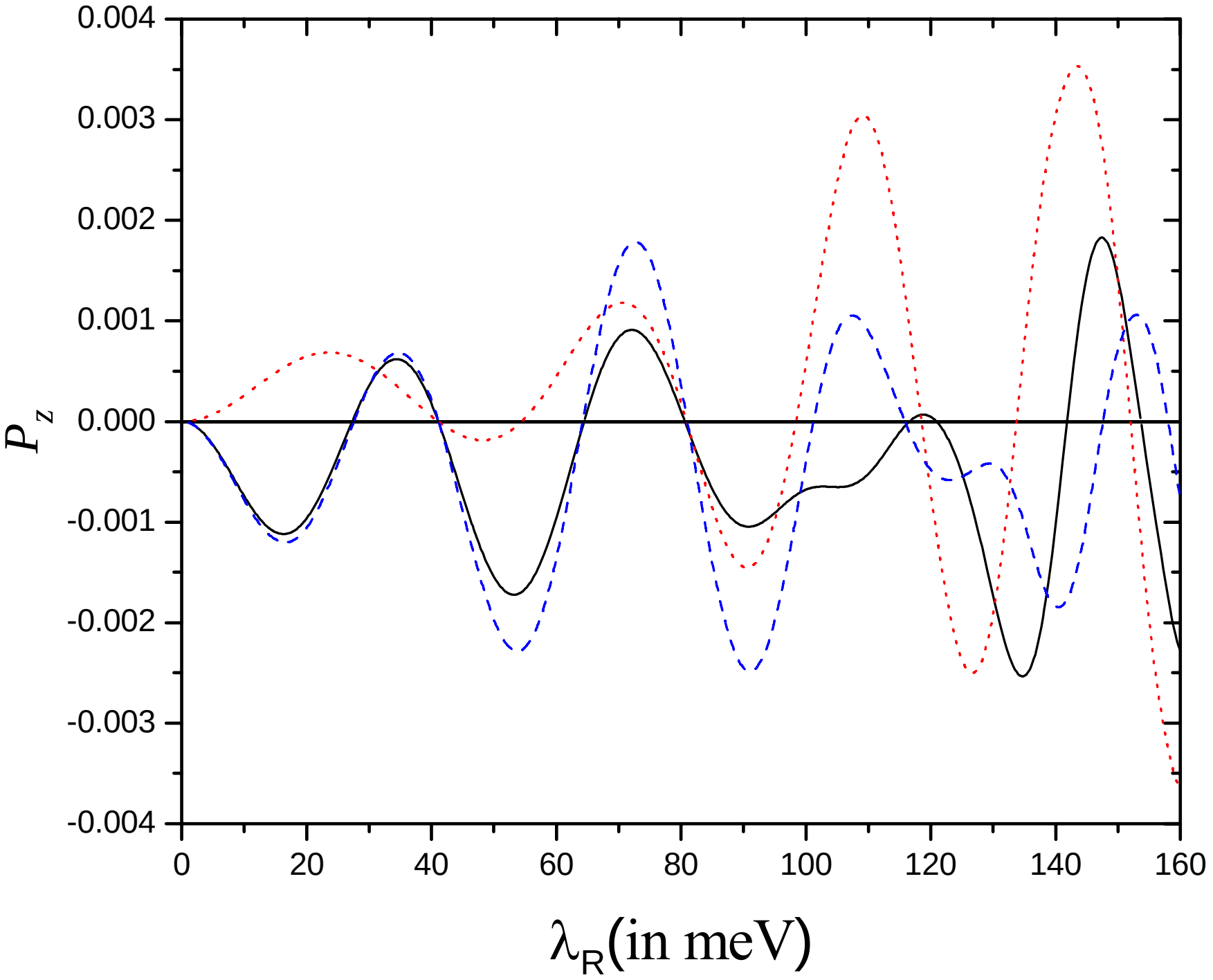

Fig.8a

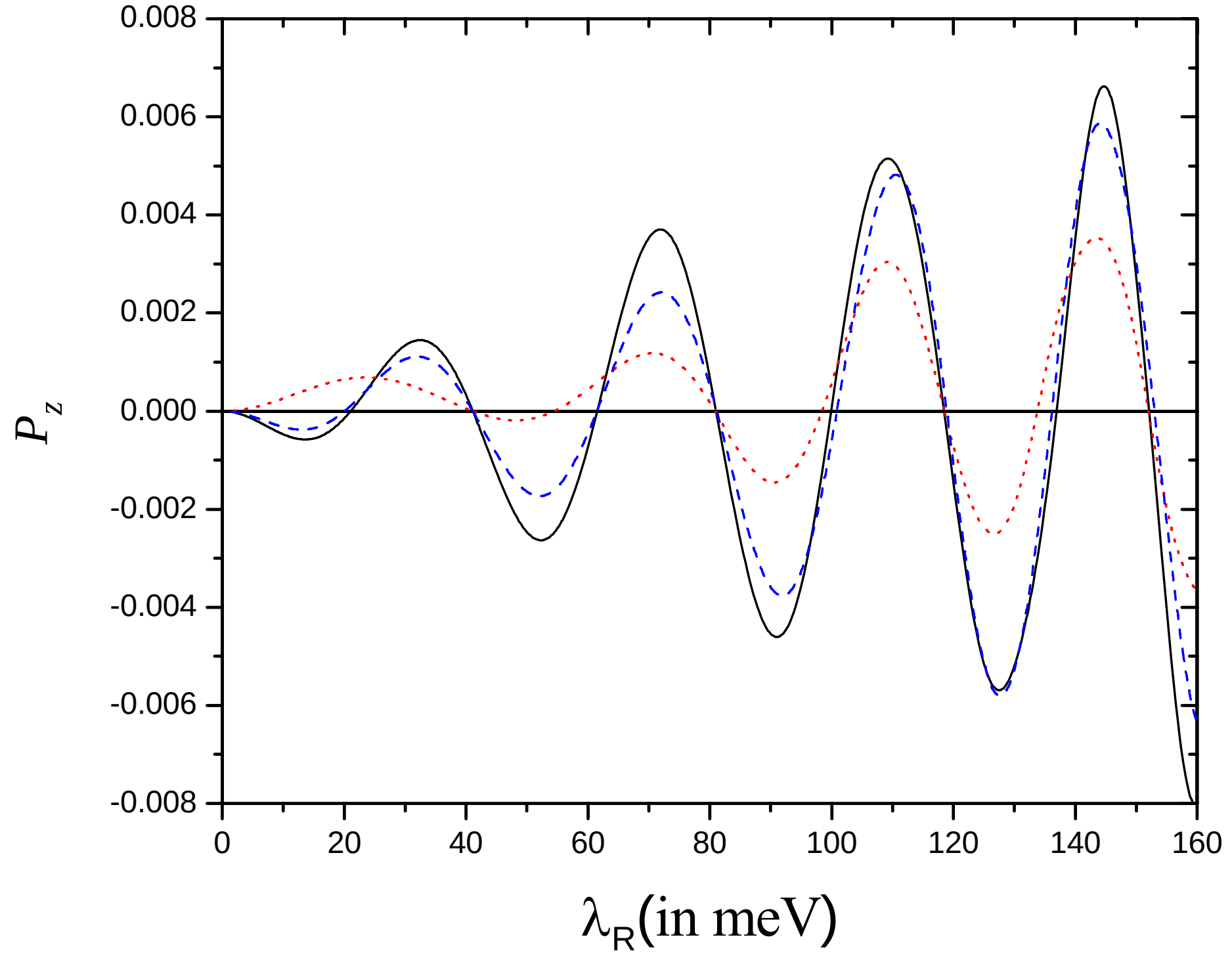

Fig.8b

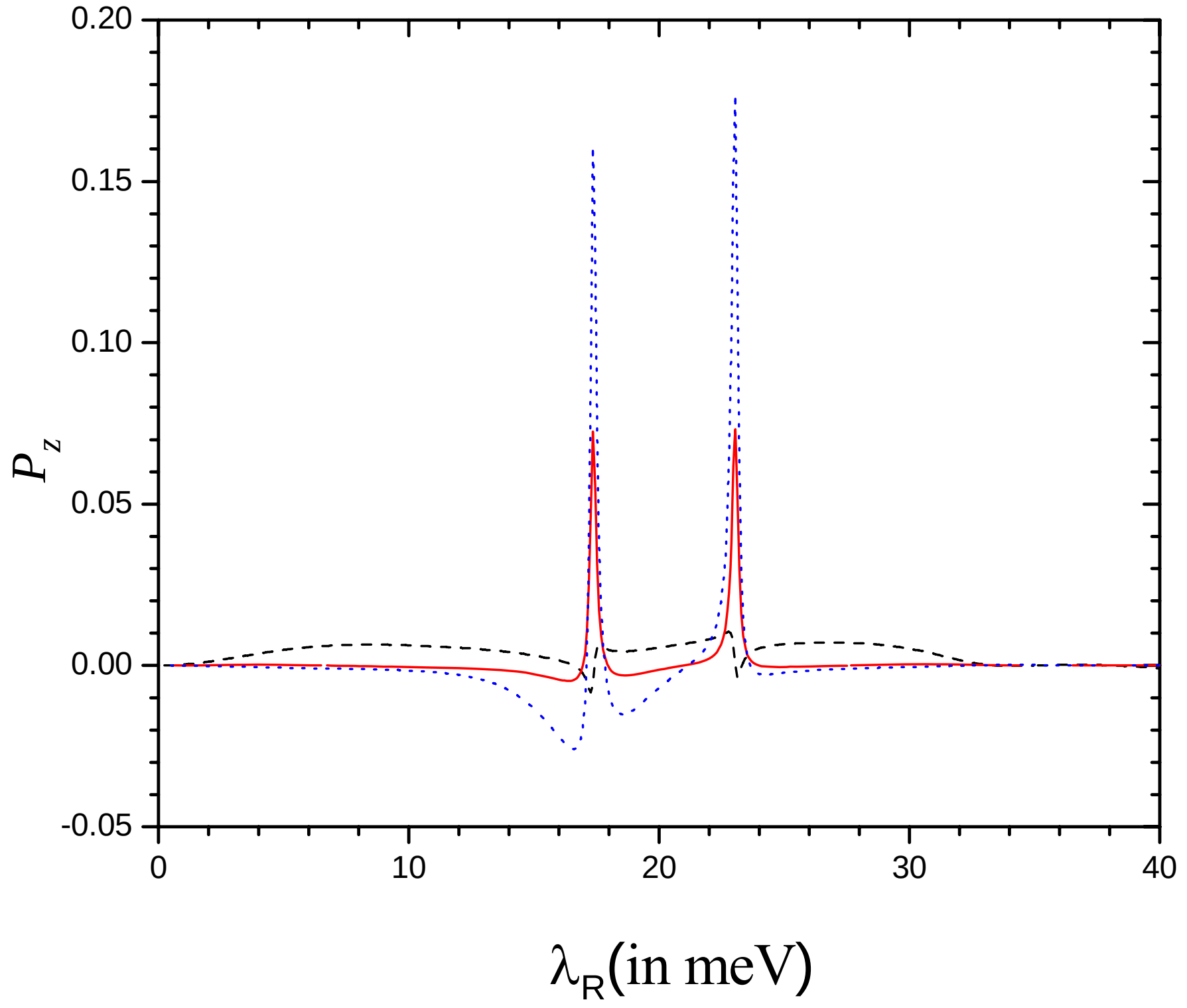

Fig.9a

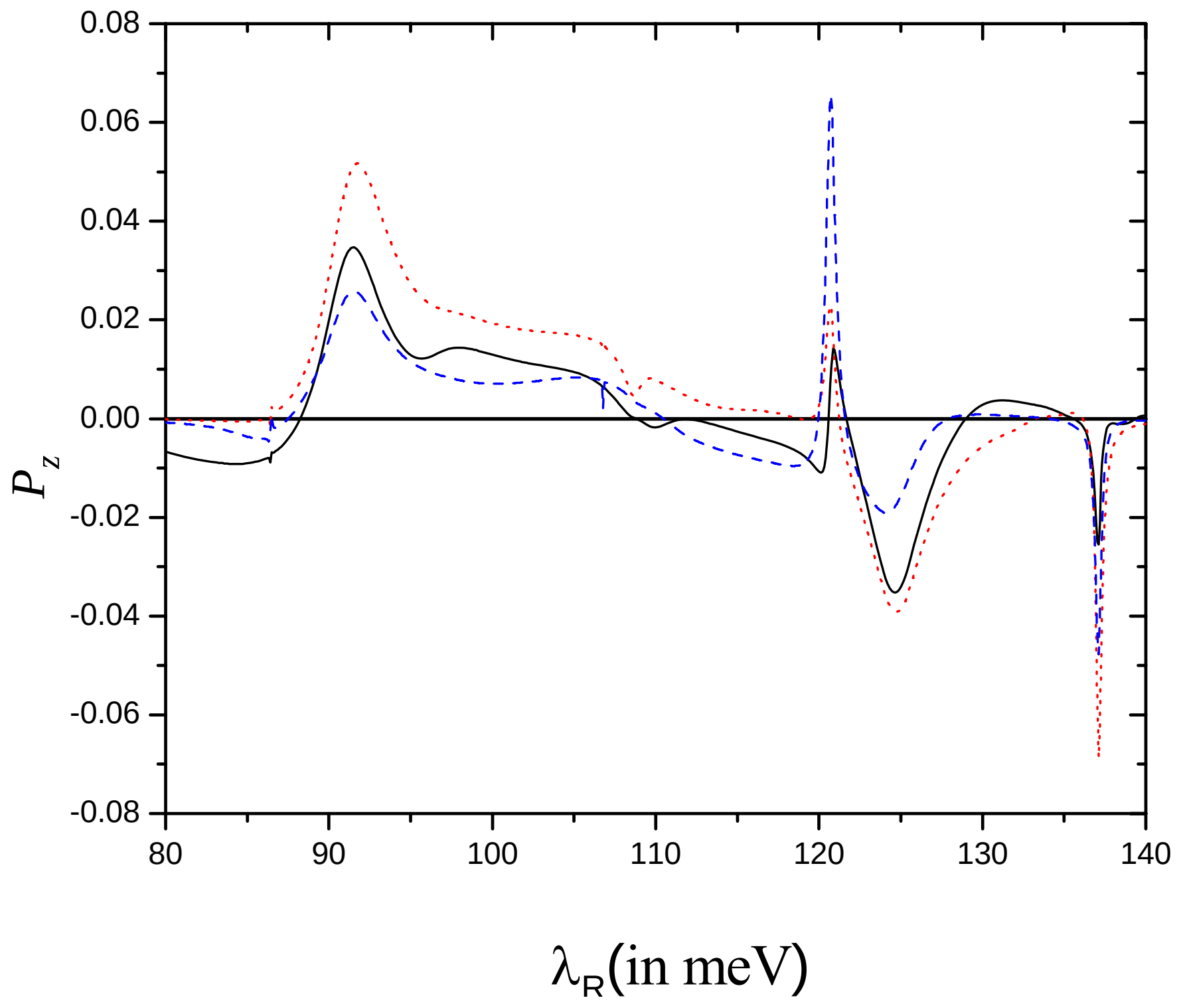

Fig.9b